# Generalized super-resolution 4D Flow MRI - using ensemble learning to extend across the cardiovascular system

Leon Ericsson, Adam Hjalmarsson, Muhammad Usman Akbar, Edward Ferdian, Mia Bonini, Brandon Hardy, Jonas Schollenberger, Maria Aristova, Patrick Winter, Nicholas Burris, Alexander Fyrdahl, Andreas Sigfridsson, Susanne Schnell, C. Alberto Figueroa, David Nordsletten, Alistair A. Young, and David Marlevi

***Abstract*— 4D Flow Magnetic Resonance Imaging (4D Flow MRI) is a non-invasive measurement technique capable of quantifying blood flow across the cardiovascular system. While practical use is limited by spatial resolution and image noise, incorporation of trained super-resolution (SR) networks has potential to enhance image quality post-scan. However, these efforts have predominantly been restricted to narrowly defined cardiovascular domains, with limited exploration of how SR performance extends across the cardiovascular system; a task aggravated by contrasting hemodynamic conditions apparent across the cardiovasculature. The aim of our study was to explore the generalizability of SR 4D Flow MRI using a combination of heterogeneous training sets and dedicated ensemble learning. With synthetic training data generated across three disparate domains (cardiac, aortic, cerebrovascular), varying convolutional base and ensemble learners were evaluated as a function of domain and architecture, quantifying performance on both *in-silico* and acquired *in-vivo* data from the same three domains. Results show that both bagging and stacking ensembling enhance SR performance across domains, accurately predicting high-resolution velocities from low-resolution input data *in-silico*. Likewise, optimized networks successfully recover native resolution velocities from downsampled *in-vivo* data, as well as show qualitative potential in generating denoised SR-images from clinical-level input data. In conclusion, our work presents a viable approach for generalized SR 4D Flow MRI, with ensemble learning extending utility across various clinical areas of interest.**

***Index Terms*—4D Flow MRI, cardiovascular, ensemble learning, hemodynamics, super-resolution,**

## I. Introduction

HEMODYNAMIC quantification is a central feature of contemporary cardiovascular medicine, with regional changes in blood flow, velocity, and pressure all indicative of disease onset and progression across the entire cardiovascular system [1]. Amongst a range of available techniques, time-resolved three-dimensional phase-contrast magnetic resonance imaging - more commonly *4D Flow MRI* - has emerged as one of the most promising imaging techniques, allowing for the non-invasive capture of full-field hemodynamics [2]. The impact of this technique has also been exemplified across various cardiovascular application areas, from the heart [3] and aorta [4], to the brain [5]. In clinical practice, the use of 4D Flow MRI is still limited by effective spatiotemporal resolution, with acquired voxel size being in direct trade-off with effective signal-to-noise ratio and required scan time. Further, with accurate image-based quantification of velocity [6], flow [7], and pressure [8], all directly dependent on acquired resolution there remains a definitive need for effective approaches to achieve high-resolution 4D Flow MRI in order to extend use across a wider spectrum of cardiovascular application areas.

To address the need for improved resolution, novel acquisition protocols or high-Tesla systems have been proposed [9], [10], however, are limited to pre-defined systems. Image-based computational fluid dynamics (CFD) provide an avenue for unrestricted resolution [11], however, require high-performance computational resources and well-defined model geometry and boundary conditions. Alternatively, deep learning methods have been proposed to enable super-resolution image conversion *post* acquisition, with networks developed in non-medical settings now entering the field of medical imaging. For anatomical super-resolution MRI, deep convolutional networks have been proposed across various application areas [12], [13], and novel generative adversarial [14] or attention networks [15] have also been introduced. For super-resolution 4D Flow MRI, residual networks have shown particular promise [16]–[18], and together with recent examples including unsupervised [19] or physics-informed neural networks (PINNs) [20], this all highlights the increasing interest in using deep learning to enhance the quality of clinically acquired flow data.

Despite this increasing interest, networks have almost exclusively been trained and tested on isolated, pre-defined cardiovascular compartments: a number of studies targeting cerebrovascular flow enhancement using cerebrovascular input data [17]; others using aortic input data to enhance aortic flow capture [16], [21]; and for super-resolution PINNs, retraining is so far required whenever transferring to a new

L.E. and A.H. contributed equally to the work and share the first-author position.

This work was funded in part by the European Union (ERC, MultiPRESS, 101075494). Views and opinions expressed are those of the authors and do not reflect those of the European Union or the European Research Council Executive Agency. Funding was also provided in part by NIH grant R01HL170059.

L.E., A.H., A.F., A.S., and D.M. are with Karolinska Institutet, Solna, Sweden. M.U.A. is with Linköping University, Linköping, Sweden. E.F. and A.A.Y. are with the University of Auckland, Auckland, New Zealand. M.B., B.H, N.B, C.A.F, and D.A.N. are with the University of Michigan, Ann Arbor, USA. J.S. is with the University of California San Francisco, San Francisco, CA, USA. M.A. ans S.S. are with Northwestern University, Chicago, USA. S.S. is also with the University of Greifswald, Germany. A.A.Y. is also with King's College London, London, UK. D.M. is also with Massachusetts Institute of Technology, Cambridge, USA.

anatomy [20]. In the setting of supervised networks, this is a particular constraint where performance will be directly dependent on required similarity between training and testing data. To exemplify, Ferdian et al. showed how application of an aortic network in a cerebrovascular setting yielded distinctive prediction biases, necessitating domain-specific training data whenever applied on novel domains [17]. Shit et al. [18] utilized training data from mulitple flow compartments, however, generalizability of super-resolution networks into *unseen* cardiovascular domains remains an unassessed problem, not least considering the contrasting hemodynamic conditions present across the cardiovascular system.

The issue of generalizability is, however, an area of active research. Data-centric approaches including data augmentation or cross-validation are commonly employed, and transfer learning strategies are tailored to improve performance beyond a pre-defined training domain [22]. Amongst available approaches, ensemble learning has emerged as an area of particular promise, where multiple base learners are combined in a meta approach to improve performance beyond that of any singular input network [23]. Crucially, ensemble learning has shown specific potential to improve out-of-distribution generalization through combination of heterogeneous base learners: either by varying training data, or by varying base architectures. While successfully employed for non-medical super-resolution imaging [24], ensemble strategies have yet to be explored for super-resolution 4D Flow MRI.

The aim of this study is therefore to evaluate the utility of ensemble learning in the setting of super-resolution 4D Flow MRI, focusing on the ability to generalize performance across multiple cardiovascular domains. Using the existing super-resolution network 4DFlowNet [16] as a base framework, and utilizing synthetic and clinically acquired 4D Flow MRI data from various cardiovascular compartments for training, testing, and validation, our contributions lie in *(1)* quantifying the limitations in generalizability of base learners trained on isolated cardiovascular compartments; *(2)* assessing the performance gain of various ensemble learning setups for improving super-resolution performance across disparate cardiovascular domains; and *(3)* translating utilities into a direct clinical setting, paving the way for super-resolution 4D Flow MRI in a more direct, cardiovascular practice.

## II. Methods

### A. Models and data preparation

*1) Patient-specific cardiovascular models:* To train a supervised super-resolution network, coupled sets of low and high-resolution images need to be acquired. In practice, collecting such paired data is inherently difficult, not least considering that high-resolution data suitable for training would require virtually noise-free, artifact-free input, acquired at resolutions beyond clinical routine. As an alternative, synthetic 4D Flow MRI originating from patient-specific CFD models have been successfully utilized [16]–[18], allowing for input data at unrestricted spatiotemporal sampling.

For the purpose of assessing generalizability, we utilize anatomically accurate patient-specific CFD models from three different cardiovascular compartments: the heart, the aorta, and the cerebrovasculature. These were purposely chosen to represent domains of disparate hemodynamic nature, ranging from high-velocity aortic jets to slow diastolic flows traversing narrow cerebrovascular arteries. With modelling details described in separate work [11], [16], [25], below follows a brief overview of utilized models:

**Cardiac:** Patient-specific models of the left heart including left atrium, left ventricle, and left ventricular outflow tract were utilized from four (n=4) different subjects, each with varying degrees of simulated mitral regurgitation (one grade 1, two grade 2, and one grade 4). Models based on medical input data were calibrated and simulated as described in Bonini et al. [25].

**Aortic:** Patient-specific models of the thoracic aorta were utilized from three (n=3) different subjects: one without any vascular disease; two with coarcted narrowings just distal to the left subclavian artery. Data was extracted from the aortic root to a distal part of the descending aorta. Models were identical to the ones simulated and used for super-resolution training in Ferdian et al. [16].

**Cerebrovascular:** Patient-specific models of the arterial cerebrovasculature were utilized from four (n=4) different subjects: one without any cerebrovascular disease, one with severe stenosis in the right proximal internal carotid artery (ICA); one with bilateral carotid stenosis; and one being the bilaterial stenosis case after surgial re-opening of the right proximal ICA. Models were identical to the ones used for super-resolution training in Ferdian et al. [17], with modelling details provided in preceeding work [11], [17].

Additionally, in order to assess network performance in an *unseen* domain, a fourth model compartment was also defined:

**Aortic dissection:** Patient-specific CFD modelling was performed on one (n=1) subject with a medically managed type B aortic dissection, exhibiting a primary entry and exit tear with no septal fenestrations in the thoracic segment. Imaging data was extracted for the entire thoracic type B aorta, covering the aortic root, branching into false and true lumen, and cutting the model at a distal descending end at around diaphragm level. Modelling was performed for this study, although CFD details follow equivalent steps presented in similar, previous work [26].

The aortic dissection model was purposely selected to represent not only a domain withheld from training, but a domain of highly complex hemodynamic nature.

*2) Synthetic image generation:* To allow for clinically relevant training data, nodal CFD data was converted into pairs of synthetic 4D Flow MRI using a pipeline described in Ferdian et al. [16], [17]. In brief, CFD output was sampled onto uniform voxelized image grids, with noise-free high resolution data generated at spatial samplings of dx = 0.5, 0.75, 1, and 1.5 mm isotropic, respectively. To create low resolution equivalents mimicking acquired 4D Flow MRI data, high resolution data was downsampled through appropriate $k$-space cropping along with the addition of zero-mean Gaussian noise in the complex signal. In our work, high:low resolution pairs were created at a factor of 1:2. Complementing the synthetic phase data, synthetic magnitude images were generated from the corresponding fluid region segmentations, obtained from the CFD output.

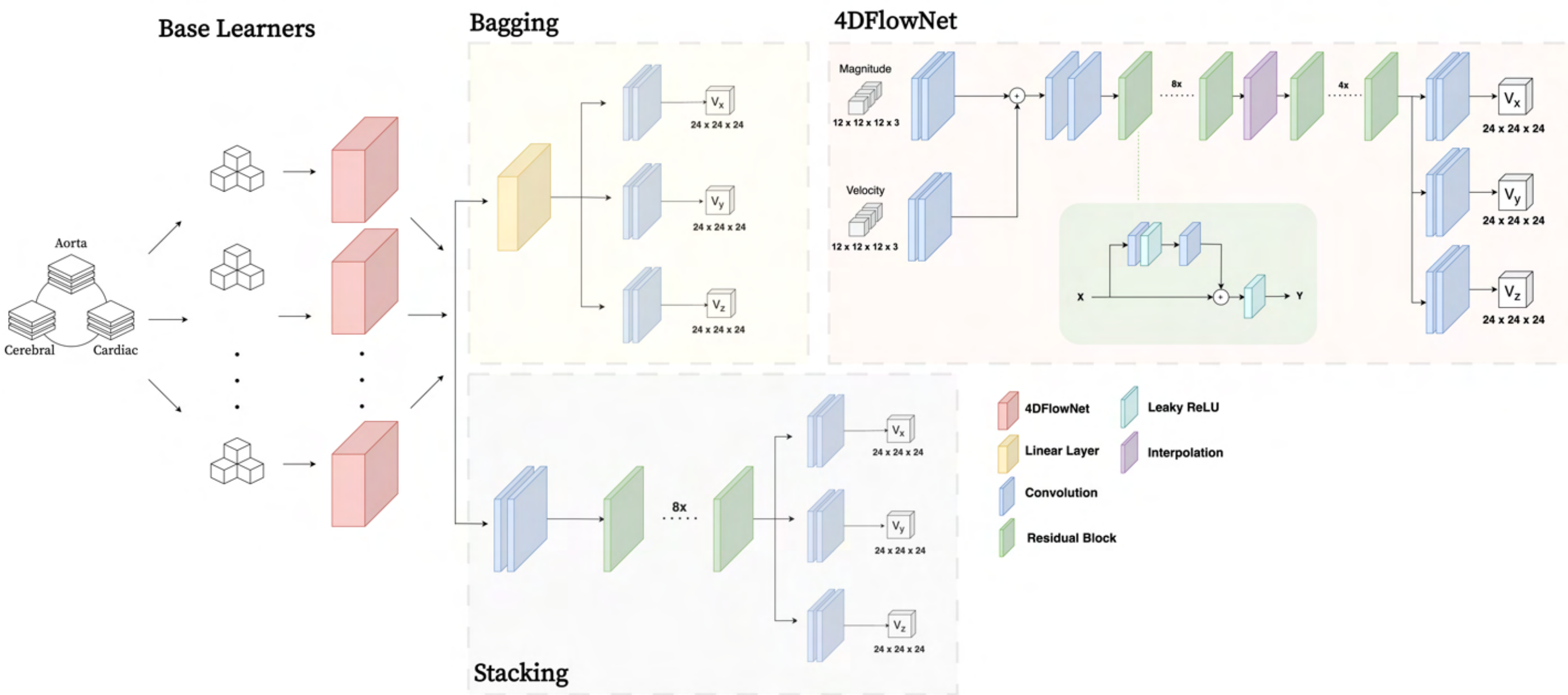

Fig. 1. Overview of the baseline network architectures, with a number of base learners drawing from pooled data before being ensembled through either bagging or stacking approaches. The base learner architecture 4DFlowNet (presented elsewhere [16] is shown in brevity on the top right).

*3) Training patches and data augmentation:* To generate a larger number of training sets, the voxelized representations were split into 3D patches of $12^3$ voxels throughout the selected field-of-view, enforcing each patch to contain a minimum of 5% non-stationary voxels. With each time frame treated independently, data heterogeneity was introduced by varying velocity encoding (VENC) across the cardiac cycle, leading to varying SNR in subsequent data patches (note that VENC was consistently kept above the maximum velocity to avoid aliasing). Patch-based data augmentation was introduced by rigid Cartesian rotation (90/180/270°) to avoid directional bias.

Through the above, a total of 13900, 21300, and 30846 patches were created for the cardiac, aortic, and cerebrovascular models, with a data split of 6:2:2 between training, validation, and testing. Note that data was partitioned model-wise rather than sample-wise to maintain integrity and independence of data during training and evaluation.

## B. Network setups

To systematically assess the impact of ensemble learning on super-resolution performance, a variety of network setups were evaluated:

*1) Baseline super-resolution network:* As a basis for comparison, the residual network 4DFlowNet [16] served as a baseline framework (see architecture in Figure 1, top right). With the network previously published and validated across various isolated domains [16], [17], it utilizes two core input paths including 3D image patches of the assessed anatomy (magnitude) and velocity (phase) for all Cartesian velocity directions. Once fed into the network, data passes through stacked convolutional and residual blocks including a core upsampling layer, before generating output in the form of super-resolved velocity patches in each Cartesian velocity direction ($v_x$, $v_y$, $v_z$).

*2) Isolated models:* To serve as a baseline for how networks trained on *isolated* cardiovascular domain perform, three 4DFlowNet networks were trained with data coming from the compartments described in Section II-A.1. This resulted in so called *isolated networks* trained only on cardiac data (4DFlowNet-Cardiac), aortic data (4DFlowNet-Aorta), or cerebrovascular data (4DFlowNet-Cerebro).

*3) Combined baseline model:* Advancing from the isolated models a *combined baseline* model was created, maintaining the 4DFlowNet architecture; however, merging datasets from all models into one. To facilitate for imbalance between compartment data, a loss function weighting scheme was introduced, balancing compartment influence on a per-batch level (see Section II-B.5).

*4) Ensemble models:* Moving beyond input data variations, two general ensemble learning approaches were explored (see Figure 1):

**Bagging:** Being one of the most common ensemble strategies, bagging consists in fitting several base models on different bootstrap samples, before aggregating them. Here, bagging was implemented using singular 4DFlowNet models as base models, with training samples drawn randomly from the available training data. Throughout, replacement sampling was allowed with base learner sample size $N$ equal to that of the original training set. A soft voting ensemble was utilized, invoking average weighting of single models in fusion prediction.

**Stacking:** Representing a second family of ensemble approaches, stacking uses a trained meta-learner as fusion of input base models. Base learners are again represented by singular 4DFlowNet models trained on sub-samples of all available training data. For the fusion meta-learner, we employed a single 8-layer convolutional feed-forward network,

TABLE I

ACQUISITION PARAMETERS FOR THE UTILIZED *in-vivo* DATA.

| | Thoracic | Cerebrovascular |
|---|---|---|
| Acquisition system | Siemens Sola | Siemens Skyra |
| Field strength [T] | 1.5 | 3 |
| Spatial resolution [mm] | 3 | 0.98 |
| Temporal resolution [ms] | 55 | 42-86 |
| Velocity encoding [cm/s] | 150 | 120 / 60 |
| TR / TE [ms] | 4.1 / 6.3 | 5.7-6.6 / 3.1-4.4 |
| Flip angle [°] | 15 | 15 |
| Gating | Retrospective ECG | Prospective ECG |
| Acceleration | Compressed sensing | *k-t* GRAPPA |
| Acceleration factor | 7.6 | 5 |

[1] TR: repetition time; TE: echo time; ECG: electrocardiogram.

with input and output identical to that of 4DFlowNet.

*5) Loss function:* The optimization target was defined by a velocity data matching term, $l_{MSE}$, given as:

$$l_{MSE} = \frac{1}{N}\sum_{i=1}^{N} \Delta v_x^2 + \Delta v_y^2 + \Delta v_z^2, \quad (1)$$

with $N$ being the total number of voxels in a given patch. To compensate for fluid/non-fluid imbalances, the loss function was split as per:

$$l_{total} = w_c(l_{MSE-fluid} + l_{MSE-non-fluid}) + \lambda \sum_{i=1}^{N} w_i^2, \quad (2)$$

with $\lambda = 5 \cdot 10^{-}7$ introduced on network weights $w_i$. $w_c$ was introduced as a compartment weight, compensating for imbalances between different training compartments by:

$$w_c = \frac{N_c}{S_c \sum_{i=1}^{N_c} K_i} \quad (3)$$

with $N_c$ the number of compartments, $S_c$ the number of samples of compartment $i$ in the assessed batch, and $K_i = \frac{1}{S_i}$.

*6) Training:* All networks were implemented using Tensorflow 2.6.0 [27] with a Keras backend [28]. The Adam optimizer was used with an initial learning rate of $10^{-4}$ and a learning rate decay of $\sqrt{2}$. Training was performed on a two NVIDIA A100 Tensor Core GPUs. With base and meta models trained for 60 and 80 epochs, respectively, this rendered a total training time of 10-15 hours for the non-ensemble, and 20-25 hours hours for the ensemble networks, respectively. Complete setup and trained weights are publicly available at `https://github.com/LeonEricsson/Ensemble4DFlowNet`.

## C. Performance evaluation

*1) Parametric in-silico validation and quantitative accuracy assessment:* To validate network performance, synthetic 4D Flow MRI data from Section II-A.1 was utilized comparing high-resolution velocities to super-resolved equivalents. Focusing on domain generalization, performance was consistently evaluated on cardiac, aortic, and cerebrovascular test cases along with overall average performance. With evaluation metrics defined in Section II-C.4, below follows a brief overview of the parametric assessments performed:

**Baseline vs. ensemble:** To provide an estimate of ensemble potential, ensemble models were initially compared against isolated and combined baseline models. Serving as a first benchmark for ensemble performance, evaluation was performed for bagging and stacking networks consisting of two homogeneous base learners.

**Number of base learners:** To assess how ensemble performance scaled with the number of input base models, ensemble models created from an increasing number of base learners were evaluated (ranging from 2 to 12).

**Compartmentalized vs. Non-compartmentalized:** To quantify how variations in base learning training data influenced performance, ensemble networks consisting of base learners sampling from a single (compartmentalized) vs. a pooled (non-compartmentalized) domain of training data were compared. Models were defined with three homogeneous base learners.

**Architectural heterogeneity:** To assess how heterogeneity in base learner architecture influenced performance, bagging and stacking models built from three homogeneous base learners were compared to models built on three heterogeneous base learners, where heterogeneity was introduced by replacing residual blocks with corresponding dense or cross stage partial blocks (similar to [21]).

*2) Quantifying generalizability into an unseen domain:* Seeking to quantify network generalizability in out-of-domain settings, ensemble networks were also evaluated on synthetic 4D Flow MRI from the *unseen* aortic dissection domain. Consistently, performance of the best performing networks from Section Section II-C.1 were compared against isolated and combined baseline models.

*3) In-vivo verification and clinical potential:* To translate the *in-silico* results into an *in-vivo* setting, network performance was evaluated on 4D Flow MRI acquired with research sequences. Data was retrospectively assembled from both thoracic (n=5) and cerebrovascular (n=5) subjects, respectively, with specific scan parameters provided in Table I. All clinical acquisitions followed institutional review board (IRB) approval, with patients referred for MRI either based on clinical indication (thoracic) or research-based study inclusion (cerebrovascular).

In lack of high-resolution reference data, we opted for *downsampling* acquired clinical data, assessing how super-resolution networks can recover initial native resolution.For this, clinical data was downsampled by a factor of two through $k$-space truncation (identical to Section II-A.2). Using our proposed baseline and ensemble networks, recovered super-resolution velocity fields were compared to the natively acquired input data, evaluating performance within left ventricular, aortic, or cerebrovascular flow domains, respectively.

*4) Evaluation metrics:* To measure network performance, relative speed error, $RE$, was defined as:

$$RE = \frac{1}{N}\sum_{i=1}^{N} \tanh\left(\frac{\|\mathbf{V}' - \mathbf{V}\|_2}{\|\mathbf{V}\|_2 + \epsilon}\right), \quad (4)$$

with $\mathbf{V}$ and $\mathbf{V}'$ being reference and predicted velocities, and with $\epsilon = 10^{-4}$ introduced to avoid zero-division. $tanh$ was introduced to mitigate over-penalizing low velocities.

Beyond the relative metric above, root mean square errors (RMSE) were estimated across the entire fluid and non-fluid

TABLE II

ESTIMATED EVALUATION METRICS ACROSS ISOLATED, COMBINED BASELINE AND ENSEMBLE MODELS WITH TWO BASE LEARNERS EACH.

| Metric | Model | Aorta | Cerebral | Cardiac | Average |
|---|---|---|---|---|---|
| RE ↓ | 4DFlowNet-Aorta | 12.36% | 34.46% | 36.99% | 27.94% |
| | 4DFlowNet-Cerebral | 49.65% | 29.58% | 69.81% | 49.68% |
| | 4DFlowNet-Cardiac | 36.22% | 37.15% | 33.02% | 35.47% |
| | Baseline Combined | 10.20% | 27.51% | 31.25% | 22.99% |
| | Bagging-2 | 10.62% | 29.54% | **30.73**% | 23.63% |
| | Stacking-2 | **10.07**% | **24.46**% | 32.20% | **22.25**% |
| RMSE ↓ | 4DFlowNet-Aorta | (1.58, 0.53, 0.73) | (1.65, 1.58, 1.62) | (2.12, 1.59, 1.98) | (1.78, 1.23, 1.44) |
| | 4DFlowNet-Cerebral | (34.01, 35.74, 30.80) | (1.04, 0.90, 0.94) | (38.18, 40.06, 35.35) | (24.41, 25.57, 22.36) |
| | 4DFlowNet-Cardiac | (4.74, 1.98, 2.22) | (1.76, 1.63, 1.75) | (2.34, 1.54, 2.13) | (2.95, 1.72, 2.03) |
| | Baseline Combined | (**1.22**, 0.45, 0.60) | (1.05, 0.83, 0.91) | (2.23, **1.43**, 1.94) | (1.50, **0.90**, 1.15) |
| | Bagging-2 | (1.42, 0.50, 0.65) | (1.07, 0.92, 0.98) | (2.33, 1.50, 2.00) | (1.61, 0.97, 1.21) |
| | Stacking-2 | (1.29, **0.40**, **0.53**) | (**0.81**, **0.80**, **0.75**) | (**2.09**, 1.62, **1.87**) | (**1.39**, 0.94, **1.05**) |
| $k$ | 4DFlowNet-Aorta | (0.961, 0.969, 1.012) | (0.822, 0.841, 0.856) | (0.862, 0.816, 0.761) | (0.882, 0.875, 0.876) |
| | 4DFlowNet-Cerebral | (0.560, 0.578, 0.587) | (0.902, 0.890, 0.912) | (0.770, 0.668, 0.672) | (0.744, 0.712, 0.724) |
| | 4DFlowNet-Cardiac | (0.610, 0.371, 0.572) | (0.834, 0.758, 0.744) | (0.842, 0.853, 0.726) | (0.762, 0.661, 0.681) |
| | Combined | (0.994, 0.964, 0.948) | (0.917, 0.927, 0.899) | (0.882, 0.863, 0.782) | (0.931, 0.918, 0.876) |
| | Bagging-2 | (**1.004**, 0.960, 0.984) | (0.916, 0.900, 0.889) | (**0.898**, **0.872**, **0.843**) | (**0.939**, 0.911, 0.905) |
| | Stacking-2 | (0.968, **0.975**, **1.003**) | (**0.923**, **0.933**, **0.933**) | (0.850, 0.856, 0.805) | (0.914, **0.921**, **0.914**) |
| $R^2$ ↑ | 4DFlowNet-Aorta | (0.978, 0.967, 0.974) | (0.881, 0.865, 0.852) | (0.858, 0.834, 0.762) | (0.906, 0.889, 0.863) |
| | 4DFlowNet-Cerebral | (0.705, 0.492, 0.586) | (0.881, 0.895, 0.880) | (0.405, 0.226, 0.256) | (0.664, 0.538, 0.574) |
| | 4DFlowNet-Cardiac | (0.929, 0.533, 0.831) | (0.861, 0.854, 0.821) | (0.847, 0.868, 0.764) | (0.879, 0.752, 0.805) |
| | Combined | (**0.987**, 0.977, 0.981) | (0.891, 0.919, 0.889) | (0.874, **0.893**, 0.794) | (0.917, **0.930**, 0.888) |
| | Bagging-2 | (0.984, 0.972, 0.979) | (0.886, 0.900, 0.881) | (0.874, 0.875, 0.798) | (0.915, 0.916, 0.886) |
| | Stacking-2 | (0.985, **0.981**, **0.985**) | (**0.938**, **0.920**, **0.925**) | (**0.884**, 0.847, **0.813**) | (**0.936**, 0.916, **0.908**) |

Each metric's best value is highlighted in **bold**. Arrows indicate direction of improvement. RMSE given in cm/s. RMSE, $k$ and $R^2$ given by ($v_x$,$v_y$,$v_z$).

domain. To quantify possible estimation bias, linear regression analysis was performed for all super-resolved networks, defining linear regression slopes, $k$, and coefficient of determination, $R^2$, for each Cartesian velocity direction, respectively.

## III. RESULTS

### A. Parametric in-silico validation and quantitative accuracy assessment

*1) Baseline vs. ensemble:* Qualitative comparison between isolated, combined baseline, and two ensemble models is presented in Figure 2. As apparent, distinct noise reduction is achieved by virtually all networks, albeit with visual artifacts when transferring isolated base models into unseen domains.

Moving into quantitative estimations, Table II presents summarized error metrics. Overall, isolated models exhibit optimal performance in the domain in which they had been trained, with poor translation into unseen domains. The combined baseline model showed apparent improvement as compared to the isolated models across all domains, with a relative error decrease of 1.77, 2.16, and 2.07% in the cardiac, aortic, and cerebrovascular domains, respectively. Underestimation bias was also mitigated by the combined baseline model, with

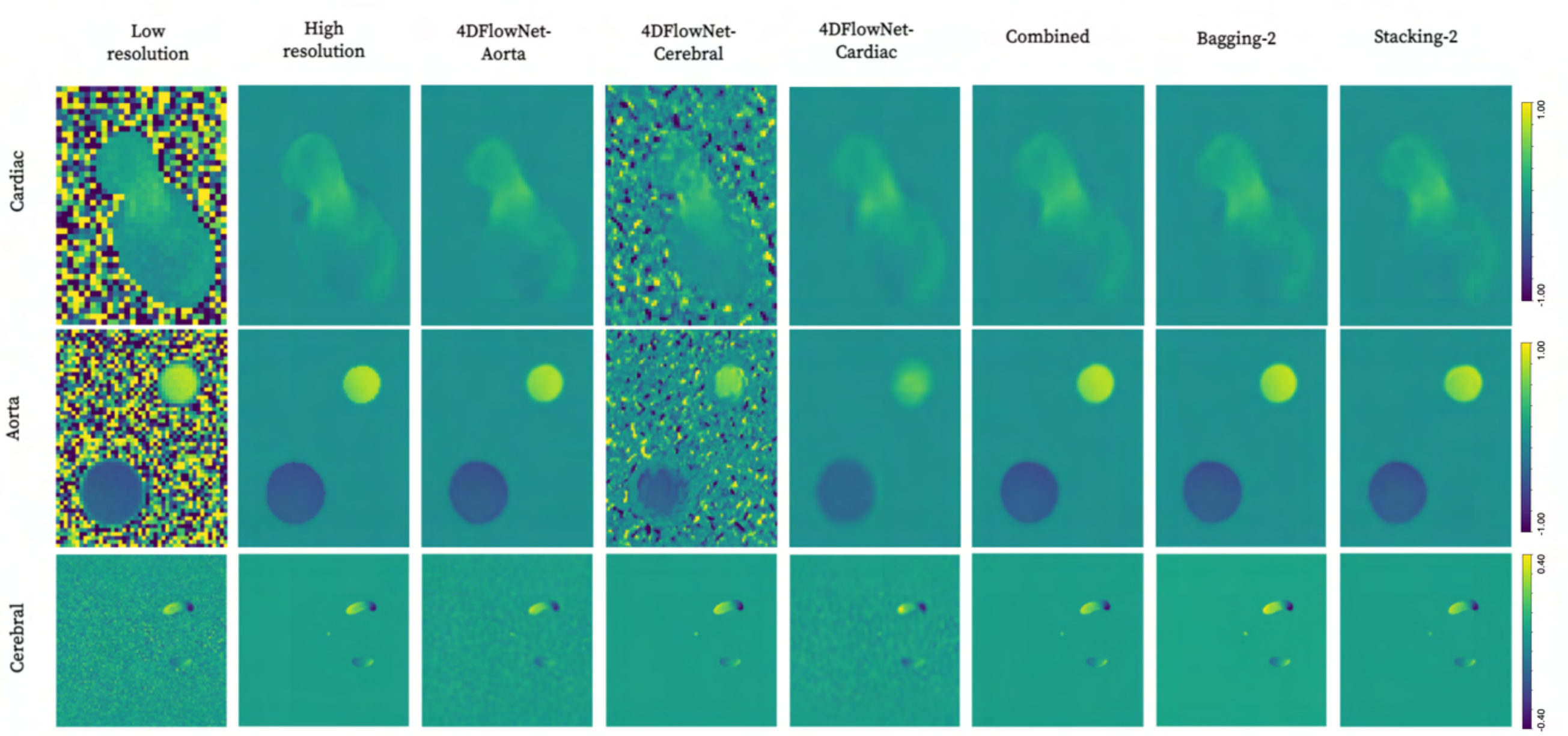

Fig. 2. Qualitative comparison of super-resolution performance across cardiac (2 mm), aortic (2 mm), and cerebrovascular (1 mm) domains using isolated, combined baseline, and first-attempt ensemble methods with two input base learners each.

TABLE III
EVALUATION METRICS FOR ENSEMBLE METHOD PERMUTATIONS INCLUDING NUMBER OF BASE LEARNERS (TOP PART), COMPARTMENTALIZED VS. NON-COMPARTMENTALIZED BASE LEARNERS (MIDDLE PART), AND BASE LEARNERS OF VARYING ARCHITECTURES (BOTTOM PART)

| | Metric | Model | Aorta | Cerebral | Cardiac | Average |
|---|---|---|---|---|---|---|
| Number of base learners | RE ↓ | Bagging-4 | 9.89% | 28.19% | 29.65% | 22.57% |
| | | Bagging-8 | 9.71% | 27.67% | 29.20% | 22.19% |
| | | Bagging-12 | **9.69%** | 27.22% | **29.13%** | **22.01%** |
| | | Stacking-4 | 10.22% | **23.97%** | 32.24% | 22.14% |
| | | Stacking-8 | 10.81% | 25.18% | 33.11% | 23.03% |
| | | Stacking-12 | 10.77% | 24.90% | 35.81% | 23.83% |
| | RMSE ↓ | Bagging-4 | (1.46, 0.47, 0.61) | (1.03, 0.86, 0.91) | (2.23, 1.45, 1.99) | (1.57, 0.93, 1.17) |
| | | Bagging-8 | (1.41, 0.44, 0.60) | (0.99, 0.85, 0.88) | (2.15, 1.40, 1.86) | (1.52, 0.90, 1.11) |
| | | Bagging-12 | (1.41, 0.44, 0.59) | (0.98, 0.83, 0.88) | (2.13, **1.38**, 1.85) | (1.51, **0.88**, 1.11) |
| | | Stacking-4 | (**1.28**, **0.42**, **0.48**) | (**0.80**, **0.76**, 0.77) | (**2.09**, 1.56, **1.84**) | (**1.39**, 0.91, **1.03**) |
| | | Stacking-8 | (1.39, 0.44, 0.59) | (0.84, 0.78, 0.78) | (2.16, 1.61, 1.97) | (1.46, 0.94, 1.11) |
| | | Stacking-12 | (1.40, 0.52, 0.57) | (0.84, **0.76**, **0.75**) | (2.10, 1.69, 1.96) | (1.44, 0.99, 1.09) |
| | $k$ | Bagging-4 | (**0.996**, **0.972**, 0.973) | (0.908, 0.901, 0.897) | (**0.890**, **0.872**, **0.825**) | (**0.931**, **0.915**, 0.898) |
| | | Bagging-8 | (0.985, 0.968, **0.979**) | (0.900, 0.899, 0.899) | (0.882, 0.868, 0.821) | (0.922, 0.912, 0.900) |
| | | Bagging-12 | (0.986, 0.964, **0.979**) | (0.899, 0.899, 0.900) | (0.883, 0.867, 0.823) | (0.923, 0.910, **0.901**) |
| | | Stacking-4 | (0.974, 0.959, 0.970) | (0.918, 0.899, 0.924) | (0.836, 0.827, 0.775) | (0.909, 0.895, 0.890) |
| | | Stacking-8 | (0.972, 0.958, 0.947) | (0.925, **0.903**, 0.906) | (0.827, 0.789, 0.711) | (0.908, 0.883, 0.855) |
| | | Stacking-12 | (0.981, 0.954, 0.969 | (**0.937**, 0.901, **0.947**) | (0.823, 0.768, 0.745) | (0.914, 0.874, 0.887) |
| | $R^2$ ↑ | Bagging-4 | (0.983, 0.975, 0.981) | (0.891, 0.910, 0.896) | (0.879, 0.884, 0.790) | (0.918, 0.923, 0.889) |
| | | Bagging-8 | (0.983, 0.977, 0.981) | (0.897, 0.911, 0.899) | (0.885, 0.893, 0.815) | (0.922, 0.927, 0.898) |
| | | Bagging-12 | (0.983, 0.977, 0.981) | (0.901, 0.915, 0.902) | (**0.887**, **0.897**, **0.817**) | (0.924, **0.930**, 0.900) |
| | | Stacking-4 | (**0.984**, **0.981**, **0.987**) | (**0.941**, **0.925**, **0.921**) | (0.881, 0.854, 0.804) | (**0.935**, 0.920, **0.904**) |
| | | Stacking-8 | (0.981, 0.976, 0.982) | (0.934, 0.913, 0.917) | (0.861, 0.837, 0.782) | (0.925, 0.909, 0.894) |
| | | Stacking-12 | (0.980, 0.969, 0.981) | (0.931, 0.919, 0.920) | (0.861, 0.816, 0.780) | (0.924, 0.901, 0.894) |
| | Metric | Model | Aorta | Cerebral | Cardiac | Average |
| Compart. vs. Non-compart. | RE ↓ | Bagging Comp-3 | 30.54% | 30.56% | 43.93% | 35.01% |
| | | Bagging-3 | 9.86% | 27.95% | **29.77%** | 22.52% |
| | | Stacking Comp-3 | 11.23% | 24.92% | 35.56% | 23.90% |
| | | Stacking-3 | **9.45%** | **24.14%** | 31.36% | **21.65%** |
| | RMSE ↓ | Bagging Comp-3 | (11.66, 11.90, 10.28) | (1.26, 1.77, 1.22) | (12.81, 13.38, 11.87) | (8.58, 8.81, 7.79) |
| | | Bagging-3 | (1.38, 0.47, 0.62) | (1.01, 0.86, 0.90) | (2.25, **1.46**, 1.96) | (1.55, 0.93, 1.16) |
| | | Stacking Comp-3 | (1.41, **0.45**, 0.58) | (0.82, **0.72**, **0.77**) | (2.21, 1.65, **1.95**) | (1.48, 0.94, 1.10) |
| | | Stacking-3 | (**1.21**, **0.39**, 0.49) | (**0.79**, 0.79, **0.77**) | (2.11, 1.50, 1.98) | (**1.37**, **0.89**, **1.08**) |
| | $k$ | Bagging Comp-3 | (0.709, 0.638, 0.723) | (0.851, 0.828, 0.836) | (0.825, 0.779, 0.720) | (0.795, 0.748, 0.760) |
| | | Bagging-3 | (**0.995**, 0.965, 0.970) | (0.916, 0.900, 0.894) | (**0.897**, **0.874**, **0.830**) | (**0.936**, 0.913, 0.898) |
| | | Stacking Comp-3 | (0.982, **1.006**, **1.005**) | (0.923, **0.920**, 0.931) | (0.845, 0.825, 0.785) | (0.917, **0.917**, **0.907**) |
| | | Stacking-3 | (0.971, 0.977, 0.987) | (**0.932**, 0.902, **0.943**) | (0.857, 0.844, 0.770) | (0.920, 0.908, 0.900) |
| | $R^2$ ↑ | Bagging Comp-3 | (0.954, 0.893, 0.927) | (0.904, 0.905, 0.892) | (0.807, 0.726, 0.679) | (0.888, 0.841, 0.833) |
| | | Bagging-3 | (0.984, 0.975, 0.981) | (0.897, 0.911, 0.897) | (**0.880**, **0.882**, **0.799**) | (0.920, **0.923**, 0.892) |
| | | Stacking Comp-3 | (0.982, 0.976, 0.983) | (0.935, **0.929**, 0.918) | (0.864, 0.825, 0.782) | (0.927, 0.910, **0.894**) |
| | | Stacking-3 | (**0.986**, **0.982**, **0.987**) | (**0.940**, 0.911, **0.920**) | (0.874, 0.865, 0.774) | (**0.933**, 0.919, **0.894**) |
| | Metric | Model | Aorta | Cerebral | Cardiac | Average |
| Arch. variation | RE ↓ | Bagging Blocks-3 | 10.35% | 27.94% | 31.28% | 23.19% |
| | | Stacking Blocks-3 | **9.67%** | **23.77%** | **31.02%** | **21.48%** |
| | RMSE ↓ | Bagging Blocks-3 | (1.39, 0.51, 0.61) | (1.13, 0.93, 0.90) | (2.21, 1.49, 1.92) | (1.58, 0.98, 1.14) |
| | | Stacking Blocks-3 | (**1.32**, **0.38**, **0.51**) | (**0.80**, **0.74**, **0.76**) | (**2.00**, **1.47**, **1.76**) | (**1.37**, **0.86**, **1.01**) |
| | $k$ | Bagging Blocks-3 | (**0.999**, 0.984, **0.977**) | (**0.956**, **0.947**, **0.931**) | (**0.910**, **0.883**, 0.812) | (**0.955**, **0.938**, **0.907**) |
| | | Stacking Blocks-3 | (0.980, **1.005**, 0.968) | (0.923, 0.909, 0.925) | (0.866, 0.837, **0.814**) | (0.923, 0.917, 0.902) |
| | $R^2$ ↑ | Bagging Blocks-3 | (**0.985**, 0.977, 0.980) | (0.901, 0.920, 0.903) | (0.881, **0.890**, 0.803) | (0.922, **0.929**, 0.895) |
| | | Stacking Blocks-3 | (0.983, **0.982**, **0.986**) | (**0.936**, **0.921**, **0.917**) | (**0.886**, 0.874, **0.825**) | (**0.935**, 0.926, **0.909**) |

Each metric's best value is highlighted in **bold**. Arrows indicate direction of improvement. RMSE given in cm/s. RMSE, $k$ and $R^2$ given by ($v_x$,$v_y$,$v_z$).

$k = 0.908$ in average across all velocity directions.

Further minor improvements were observed when moving into the first-approach ensemble models: stacking outperforming the combined baseline model across a majority of domains (relative error of 22.25% vs. 22.99%), whilst bagging exhibits slightly higher deviations (relative error of 23.63%). Similar indications are observed for RMSE, $k$, and $R^2$: stacking, bagging, and combined baseline model showing optimal performance across 25, 5, and 6 out of 36 assessed metrics.

*2) Parametric ensemble analysis:* Quantitative results for the parametric ensemble analysis is presented in Table III.

**Number of base learners:** Keeping all base learners identical, bagging scaled with the number of base learners with performance peaking at 12 base learners (average RE $= 22.01\%$, mean RMSE $=$ 1.17 cm/s). In contrast, stacking displays inverse behaviour, with accuracy decreasing with increasing number of homogenous base learners (RE $= 22.14\%$, given at two base learner). This holds true also for bias metrics from the linear regression analysis. Comparing the two approaches, the best bagging vs. stacking approach are seemingly interchangeable, with strong correlations and low errors observed across all domains (18 vs. 22 metrics perform better in bagging vs. stacking across all domains)

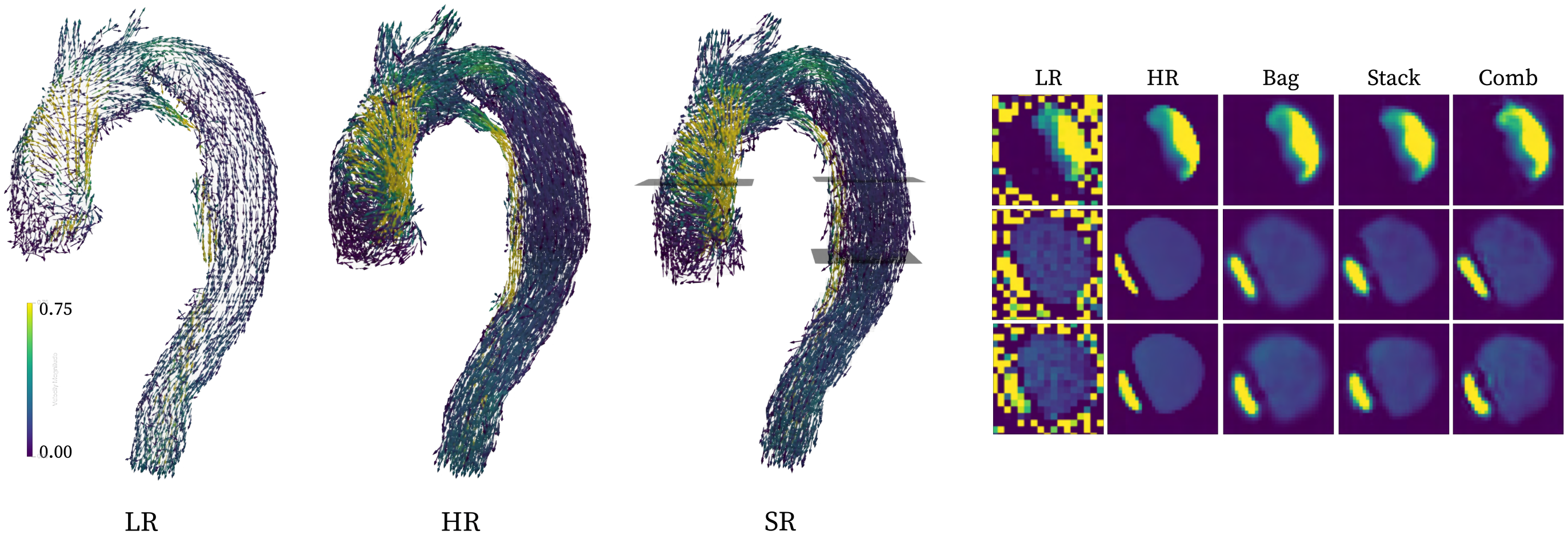

Fig. 3. Qualitative visualization of super-resolution conversion of the unseen aortic dissection domain (left, with the stacking setup representing the super-resolution conversion), along with representative cross-sections (right). All renderings are performed using calculated velocity magnitudes.

**Compartmentalized vs non-compartmentalized:** As given in the middle section of Table III, stacking is able to fuse compartmentalized base learners better than bagging, with an average relative error of 23.90% vs. 35.01%. As compared to other permutations, compartmentalized ensemble models consistently underperform as compared to non-compartmentalized equivalents. This holds across all metrics, with bagging particularly suffering from compartmentalized learners (relative errors $> 30\%$).

**Architectural heterogeneity:** The bottom part of Table III provides results for bagging and stacking containing base learners with varying architectural blocks. The given stacking permutation (*Stacking Blocks-3*) shows the best overall performance of all stacking variations (average relative error of 21.48%, average $RMSE = 1.08$ cm/s, $k = 0.933$, and $R^2 = 0.933$). Bagging on the other hand does not show the same benefit of architectural heterogeneity, where instead a maximized number of input learners (*Bagging-12*) is the model with optimal performance across all permutations.

## B. Quantifying generalizability into an unseen domain

Table IV provides evaluation metrics for the unseen aortic dissection, with Bagging-12 and Stacking Blocks-3 used as optimal ensemble models. As observed, isolated models exhibit significant difficulties translating into an unseen domain, with the cerebrovascular network having particularly poor performance (relative error $= 75.57\%$, average RMSE $=$ 60.37 cm/s). In comparison, ensemble methods exhibit high accuracy across all metrics with relative error $= 25.42$ and 24.82%, and average RMSE $= 2.63$ and 2.17 cm/s given for bagging and stacking, respectively. Concerning estimation bias, combined baseline, bagging, and stacking all show highly accurate behaviour, exhibiting high accuracy and low spread (linear regression data shown in Figure 4). Further, qualitative renderings of recovered flow features are shown in Figure 3.

TABLE IV
PREDICTION ERRORS OF ISOLATED, COMBINED AND ENSEMBLE MODELS ON THE UNSEEN AORTIC DISSECTION DATA.

| Metric | Model | Aortic dissection |
|---|---|---|
| RE ↓ | 4DFlowNet-Aorta | 34.70% |
| | 4DFlowNet-Cardiac | 40.22% |
| | 4DFlowNet-Cerebro | 75.57% |
| | 4DFlowNet-Combined | 30.53% |
| | Bagging-12 | 25.42% |
| | Stacking Blocks-3 | **24.82%** |
| RMSE ↓ | 4DFlowNet-Aorta | (2.01, 2.30, 4.07) |
| | 4DFlowNet-Cardiac | (1.86, 2.56, 5.93) |
| | 4DFlowNet-Cerebro | (54.29, 54.41, 72.42) |
| | 4DFlowNet-Combined | (2.41, 2.75, 5.19) |
| | Bagging-12 | (1.62, 1.95, 4.32) |
| | Stacking Blocks-3 | (**1.28, 1.73, 3.49**) |
| $k$ | 4DFlowNet-Aorta | (0.927, 0.819, 0.851) |
| | 4DFlowNet-Cardiac | (0.711, 0.569, 0.507) |
| | 4DFlowNet-Cerebro | (0.965, 0.692, 0.903) |
| | 4DFlowNet-Combined | (1.067, **0.994**, **0.978**) |
| | Bagging-12 | (**1.002**, 0.982, 0.958) |
| | Stacking Blocks-3 | (0.917, 0.880, 0.893) |
| $R^2$ ↑ | 4DFlowNet-Aorta | (0.824, 0.805, 0.880) |
| | 4DFlowNet-Cardiac | (0.778, 0.739, 0.779) |
| | 4DFlowNet-Cerebro | (0.254, 0.215, 0.612) |
| | 4DFlowNet-Combined | (0.836, 0.847, 0.866) |
| | Bagging-12 | (0.889, **0.904**, 0.903) |
| | Stacking Blocks-3 | (**0.905**, 0.894, **0.921**) |

Note: Each metric's best value is highlighted in **bold** font. Arrows indicate direction of improvement. RMSE given in cm/s. RMSE, $k$, and $R^2$ given by $(v_x, v_y, v_z)$

## C. In-vivo verification and clinical potential

*1) Quantitative assessment through recovery of native resolution:* Figure 5 shows exemplary *in-vivo* images, using super-resolution to recover native input resolution. Qualitatively, both ensemble networks recover high-resolution features along with background noise suppression. Behaviour also seem robust across all domains, with both large-vessel aortic and small-vessel cerebrovascular features captured.

Quantifying the above, summarized linear regressions statics are provided in Table V. Consistently, relative errors are lower using ensemble techinques, with Bagging-12 indicating optimal performance across all domains (average relative error $= 39.85\%$). Conversely, bias metrics show slight favouring of the baseline combined approach, with an average $k =$ 0.954 compared to 0.873 and 0.753 for bagging and stacking, respectively. However, regression spread is lower with ensemble techniques, with bagging exhibiting maximum specificity (average $R^2 = 0.815$ vs. 0.796 and 0.786 for combined baseline and stacking, respectively).

TABLE V

ESTIMATED EVALUATION METRICS FOR RECOVERY OF NATIVE *in-vivo* RESOLUTION ACROSS DIFFERENT CARDIOVASCULAR DOMAINS.

| Metric | Model | Aorta | Cardiac | Cerebral | Average |
|---|---|---|---|---|---|
| RE ↓ | Baseline Combined | 39.46 ± 6.81% | 45.92 ± 5.44% | 49.15 ± 4.74% | 44.84 ± 5.73% |
| | Bagging-12 | **34.11 ± 7.25%** | **42.10 ± 4.78%** | **43.34 ± 4.31%** | **39.85 ± 5.60%** |
| | Stacking Comp-3 | 36.11 ± 9.89% | 44.81 ± 5.26% | 45.35 ± 4.34% | 42.09 ± 6.94% |
| $k$ | Baseline Combined | **0.854 ± 0.156** | 1.034 ± 0.047 | **0.975 ± 0.126** | **0.954 ± 0.119** |
| | Bagging-12 | 0.803 ± 0.172 | **0.967 ± 0.042** | 0.848 ± 0.096 | 0.873 ± 0.116 |
| | Stacking Comp-3 | 0.716 ± 0.177 | 0.866 ± 0.027 | 0.677 ± 0.072 | 0.753 ± 0.111 |
| $R^2$ ↑ | Baseline Combined | 0.789 ± 0.113 | **0.877 ± 0.051** | 0.722 ± 0.057 | 0.796 ± 0.079 |
| | Bagging-12 | **0.818 ± 0.124** | **0.877 ± 0.033** | **0.749 ± 0.057** | **0.815 ± 0.081** |
| | Stacking Comp-3 | 0.807 ± 0.134 | 0.840 ± 0.031 | 0.710 ± 0.060 | 0.786 ± 0.087 |

Each metrics best value, per compartment, is highlighted in **bold**. Arrows indicate direction of improvement for each metric.

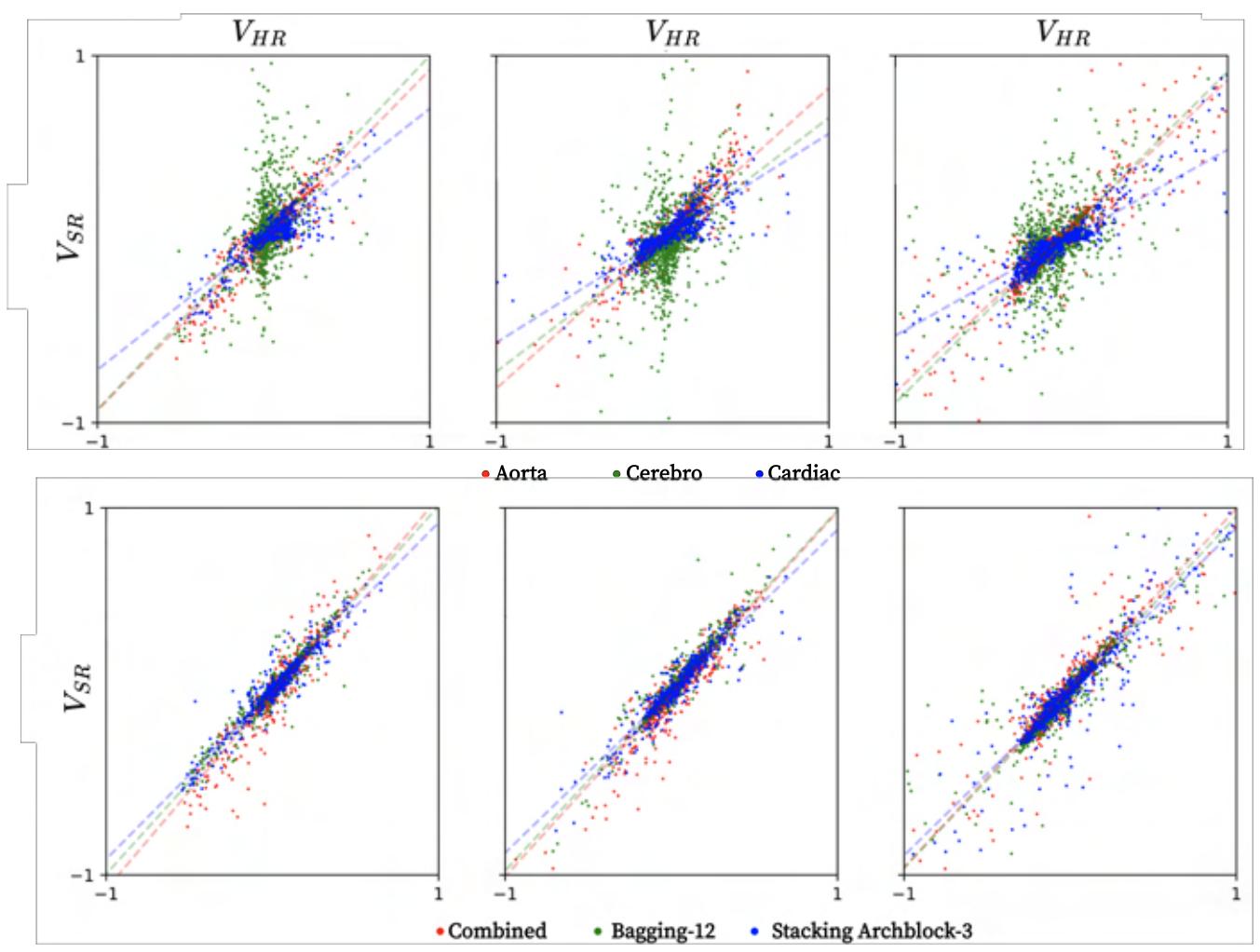

Fig. 4. Linear regression plots in the unseen aortic dissection, given for isolated baseline (top row) and combined baseline, bagging, and stacking (bottom row) learners, respectively, showing velocities in x, y, and z from left to right, all normalized to a [-1,1] range.

## IV. DISCUSSION

In this study, we have evaluated the utility of ensemble learning for super-resolution 4D Flow MRI, assessing its ability to generalize across various cardiovascular domains. As reported, ensembling along with incorporation of disparate training data distinctly improves domain generalization, with recovery of high-resolution velocities validated on both synthetic and clinical datasets across the heart, aorta, and brain. Considering the disparate hemodynamic conditions apparent across the cardiovasculature, our results thus bear particular clinical promise, opening up for generalizable super-resolution performance across domains using a single network setup.

### A. Base vs. ensemble learning

As observed across all synthetic datasets, ensemble approaches consistently outperform isolated base learners (Table II). Notably, the benefit is not only observed when moving outside an isolated model's domain-of-training, but benefits are seen even *within* the setting of an isolated learner. These results not only speak to the benefit of ensemble approaches[29] but also highlights limitations in utilizing isolated training beds with a limited number of patient sets. In our work, isolated networks are trained on ∼20'000 patches: a figure comparable to what has been previously utilized for medical super-resolution [16]–[18], however, small in contrast to non-medical equivalent. Increasing training data is a common strategy for improved performance, but here our work highlight the benefit of doing so using data from various compartments with the combined baseline model outperforming isolated learners. Adding ensemble strategies can further improve performance, enabling optimal weighting between individual learners.

The benefit of ensembling and data pooling is emphasized when transitioning into the unseen aortic dissection where all isolated models show significant errors. Poor domain generalization has been reported for networks trained on single-domain data [29], however, our results corroborate this in the setting of 4D Flow MRI. Moving into an unseen domain also highlights the benefits of combining learners, with ensemble networks improving on the combined baseline model. This is a particularly important feature in seeking generalizable performance, where data heterogeneity is observed both between domains and patients. The use of ensemble approaches thus opens for more unified analysis, super-resolving datasets at maintained accuracy in a diverse clinical reality.

### B. Parametric ensemble evaluation

In an attempt to optimize ensemble performance, a range of networks were assessed in Section III-A. Although variations were overall minor, a few notable trends can be observed:

First, the number of base learners had opposite effects on the two assessed approaches: bagging improving but stacking worsening with an increased number of base learners. For bagging, being a mere deterministic aggregation of base learners, bias and variance is typically reported to decrease with number of base learners, leading to an accuracy plateau at an empirically determined base learner density [30]. The meta-architecture of stacking, on the other hand, does not scale with base learner quantity but rather with base learner *diversity* [31]; a fact corroborated by the results on architectural variations.

Second, the use of compartmentalized base learners had a consistently detrimental effect on overall performance. The reason to this most likely lies in the underperformance of our isolated base learners, where ensemble combinations alone cannot overcome the bias exhibited by the base learners themselves. Bagging suffers particularly from compartmentalized base learners, where deterministic weight averaging renders pronounced errors across all domains. These results are contradictory to the notion that input diversity is viewed as one way of improving ensemble performance [23], [29], however, this does not necessarily cover scenarios where base learners are extended into out-of-distribution settings.

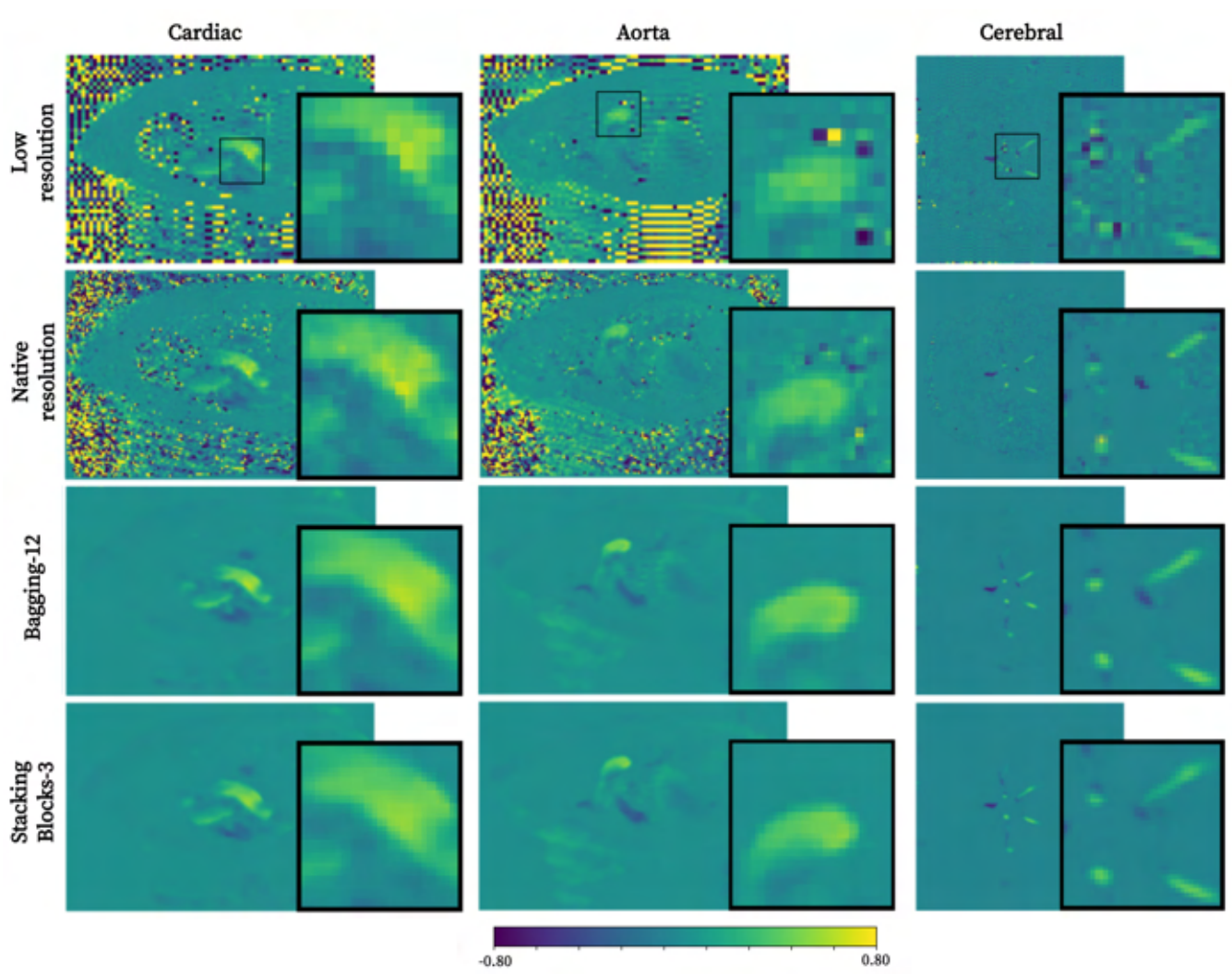

Fig. 5. Recovering *in-vivo* native resolutions using super-resolution conversion from downsampled images.

Third, architectural variations were beneficial for the stacking setup, with optimal performance given for the *Stacking Blocks-3* network. In our study, architectural variation was not achieved by replacing overall architecture, but by replacing internal layers similar to how 4DFlowNet has been altered in previous work [21]. Higher-degrees of heterogeneity could be offered by combining super-resolution networks of different core architectures, being as of yet unexplored for super-resolution 4D Flow MRI.

### C. In-vivo feasibility and clinical utility

To explore clinical translation, ensemble networks were assessed *in-vivo*, recovering native resolution from synthetically downsampled data. As reported, performance is kept stable across domains, although biases and errors are more pronounced as compared to the *in-silico* results. Here, comparisons between *in-silico* and *in-vivo* results should be viewed in light of the inherent differences between the datasets. In the *in-vivo* setting, 4DFlowNet is actually *not* trying to recover native input images directly, but rather a *de-noised* equivalent. As such, increased *in-vivo* errors do not necessarily stem from sub-optimal network performance, but also from differences between noisy native, and de-noised recovered images.

As a final note on clinical utility, it is worth highlighting that our networks are directly applicable for true super-resolution image conversion. To exemplify, Figure 6 showcases two such qualitative examples, indicating how both intracardiac vortices and cerebrovascular flow features can be resolved at beyond clinical resolution, all using a single ensemble network.

### D. Scientific contextualization

Whereas, to the best of our knowledge, no previous work have attempted ensembling techniques to super-resolve 4D Flow MRI data, or explored generalizability of super-resolution 4D Flow MRI, it is worth contrasting our results to previously published work within related spaces.

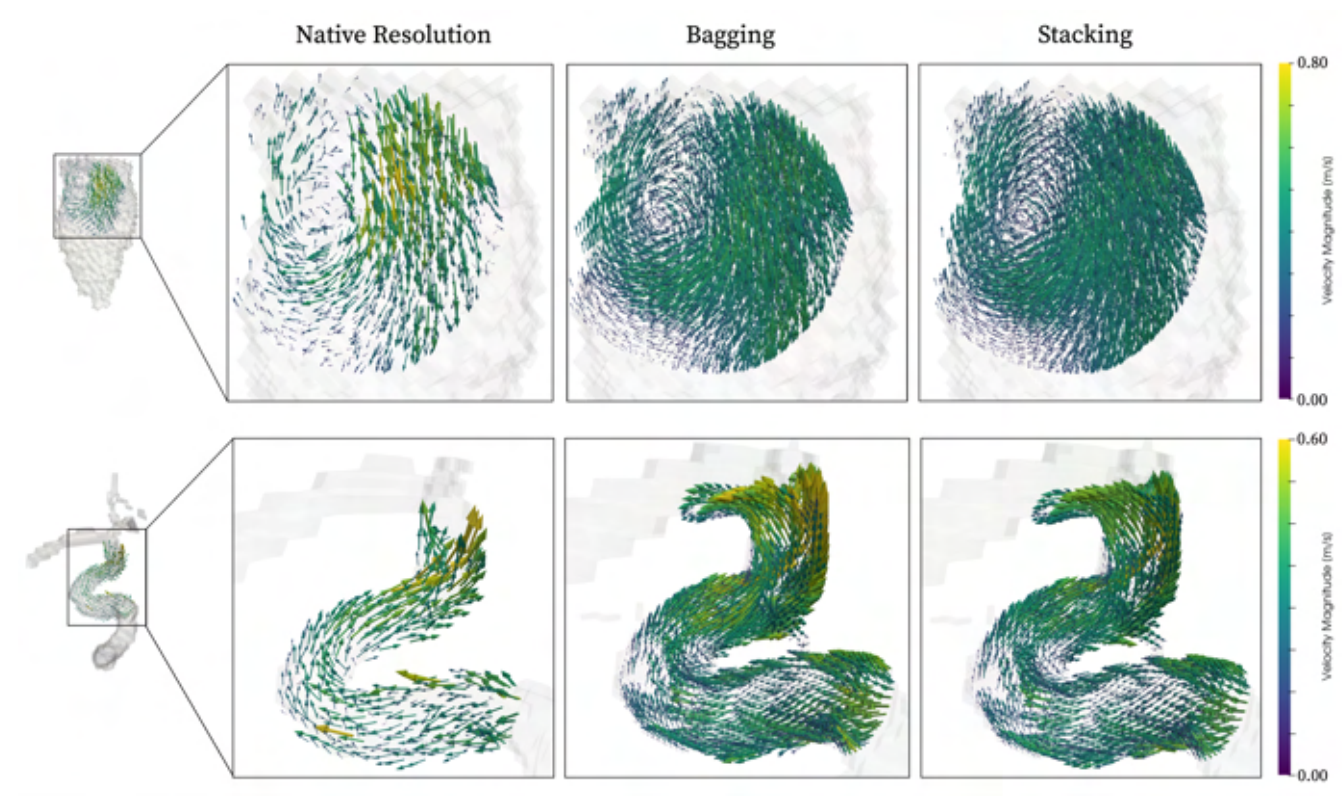

Fig. 6. Qualitative vector field rendering of clinical data upsampled by a factor two *beyond* native acquisition resolution. Examples shown for both cardiac (top) and cerebrovascular (bottom) data, for both bagging (middle) and stacking (right).

In the non-medical field, ensemble learning has been reported as one of the more promising domain generalization approaches [29]. Ju et al. evaluated bagging and stacking of residual learners for image classification, reporting incremental performance improvement in-line with our findings [32]. Similarly, Nguyen et al. [33] utilized stacking of heterogeneous learners, reporting slight improvement as compared to single base learners. For improved super-resolution generalization, examples include exploration of heterogeneous training data [34], or leveraging domain-specific image priors [35]. The latter presents an appealing approach for unifying behaviour across e.g. vendors or centres; however, for the sake of generalizability across flow domains, hemodynamic differences are inherent to the physiological nature of the observed domain.

In a medical setting, Lyu et al. [13] presented one of few examples using ensemble learning for super-resolution MRI. Using generative adversarial networks in a stacking setup they highlighted the ability to super-resolve anatomical MRI, however, focusing on a single anatomical domain. For 4D Flow MRI, Shit et al. [18] trained on both thoracic and cerebrovascular data, using transfer learning to translate *in-silico* results to *in-vivo*. Although a direct comparison is obstructed by differences in available datasets, our reported ensemble output (RMSE ~1-2 cm/s) appear non-inferior in comparison across all domains (RMSE ~2-4 cm/s). Beyond this, recent PINN work [20] promises increased super-resolution accuracy, but their utility in an ensemble setting remains to be assessed.

### E. Limitations and future work

A few limitations are worth pointing out. First, training was performed on synthetic 4D Flow MRI without inclusion of acquired *in-vivo* data. Acquiring clinical data for super-resolution purposes is difficult due to practical considerations (scan time, SNR), not least considering the notion of going *beyond* practical resolution limits. The use of $k$-space data conversion is instead purposely introduced to mitigate the effect of *in-silico*-to-*in-vivo* discrepancies, resembling the sub-sampling of an MR scanner.

Second, although tested with respect to recovery of native resolution, no *in-vivo* comparison was performed between acquired high- and acquired low-resolution data. This again comes down to the problem of acquiring paired high- and low-resolution data. The concept of recovering downsampled data has been explored by others in previous super-resolution work [18], highlighting the practicality of the approach.

For future work, a number of directions can be envisioned including exploration of diverse base learners, or incorporation of clinical training data. Efforts to integrate super-resolution algorithms as an in-line scanner utility would also greatly improve use-cases. Nevertheless, our data highlights how ensemble techniques could help generalize the use of super-resolution imaging, circumventing the need for purpose-built networks and opening for wider incorporation of super-resolution imaging in cardiovascular 4D Flow MRI work.

## V. Conclusion

In this study, we have shown how ensemble learning enables super-resolution conversion of clinically acquired 4D Flow MRI, with accurate performance generalizing across disparate flow domains. Using a combination of synthetic training data from different cardiovascular compartments, we have shown how ensemble approaches maintain accurate performance across unseen domains, as well as improve on singular base learner performance. Satisfactory recovery of native resolution *in-vivo* also highlights performance transfer into a direct patient setting, applicable across the heart, aorta, and brain.

## Acknowledgment

Computations were performed on resources provided by the National Academic Infrastructure for Supercomputing in Sweden at the National Supercomputer Centre at Linköping University (Berzelius). We also thank Ning Jin, PhD, at Siemens Medical Solutions USA, Inc., and Daniel Giese, PhD, at Siemens Healthcare GmbH, Erlangen, Germany for providing the thoracic 4D Flow research sequences.